\documentclass[prl,twocolumn,aps,showpacs]{revtex4}
\usepackage{graphicx}
\usepackage{epsfig}
\usepackage{bm,hyperref}

\def\be{\begin{equation}}
\def\ee{\end{equation}}
\def\ba{\begin{eqnarray}}
\def\ea{\end{eqnarray}}
\def\r{{\rm \bf r}}

\def\k{{\rm \bf  k}}

\begin{document}
%
%\begin{CJK*}{UTF8}{gbsn}
\title{Self-consistent Single-band Approximation for  Interacting Boson Systems}
%
%\author{Biao Wu(吴飙)}
\author{Biao Wu}
%\email{bwu@aphy.iphy.ac.cn}
\affiliation{Institute of Physics, Chinese Academy of Sciences,
Beijing 100190, China}
%
%\author{Junren Shi(施均仁)}
\author{Junren Shi}
\affiliation{Institute of Physics, Chinese Academy of Sciences,
Beijing 100190, China}
\date{\today}
%\date{July 5th, 2006}
%
\begin{abstract}
Traditionally, the single-band approximation for interacting many-body systems 
is done with pre-determined single-particle Wannier functions, ignoring the dependence
of the Wannier function on interaction.  We show that  the single-band approximation
has to be done self-consistently to properly account the interaction effect on the 
Wannier functions.  This self-consistent single-band approximation leads to 
a   nonlinear equation for Wannier functions, which can be recast into
a set of nonlinear equations for  Bloch functions. 
These equations are simplified for two special cases,  the superfluid regime and 
deep in the Mott insulator regime.  A simple example with double-well potential is 
used to illustrate our results.
\end{abstract}
%03.75.Lm  Tunneling, Josephson effect, Bose-Einstein condensates 
          %in periodic potentials, solitons, vortices, and topological excitations
%67.80.-s 	Solid helium and related quantum crystals
%67.40.-w 	Boson degeneracy and superfluidity of 4He
%03.75.Hh	Static properties of condensates; thermodynamical, statistical, and structural properties
%71.10.Fd	Lattice fermion models (Hubbard model, etc.)
%67.80.kb	Supersolid phases on lattices
%05.50.+q	Lattice theory
%37.10.Jk	    Atoms in optical lattices
\maketitle
%\end{CJK*}
Interacting many-body systems are notoriously hard to solve. One usual tactic is 
the single-band 
approximation, where the Wannier functions\cite{Wannier1937PR} for the lowest Bloch 
band are used to reduce the system to a lattice model, such as 
Hubbard model\cite{Ashcroft1976Book,Jaksch1998PRL}.  
Traditionally, the Wannier functions used in the approximation are 
obtained either directly
from the single-particle Bloch waves or with some variational approaches\cite{Kohn1973PhysRevB,Marzari1997PRB}. 
The interaction effect on the Wannier function is completely neglected 
in all these traditional methods.
This neglect  is at odds with the fact that the interaction
evidently affects
the shape of the Wannier functions. 

Current strong interest in ultracold atoms in optical lattices has put this problem into
spotlight\cite{Bloch2008RMP}. A recent experiment with ultracold atoms
clearly demonstrated that the on-site interaction, therefore, the Wannier functions, 
depends on the occupation number\cite{Campbell08042006}.
It has also seen increased theoretical efforts to address this problem
\cite{Li2006NJP,Mueller2009,Liang2009,Schneider2009,Esry1997PRL,Masiello2005PRA}.
However, all the efforts have certain drawbacks  from a general perspective. 
In the variational approach adopted by Li {\it et al.} \cite{Li2006NJP}, there
is arbitrariness in the choice of  the  trial Wannier function. The mean-field method used 
by Liang {\it et al.}  can apply only in the superfluid regime and can not be justified to apply 
for the more interesting
superfluid-Mott insulator transition regime\cite{Liang2009}.  
The {\it ab initio}  calculation  in Ref. \cite{Schneider2009} is hard to be scaled up for
systems with large number of particles. 

We address the interaction effect on Wannier function by re-examining the single-band
approximation. We find that the interaction effect
can be taken into account automatically once  the  single-band approximation is done self-consistently.
The self-consistent single-band approximation is achieved by minimizing the ground state
energy of the system by varying  Wannier function. The minimization leads to
a nonlinear equation for the Wannier function, which depends on the ground state
of the system. Therefore, one has to solve the nonlinear equation  self-consistently with
the ground state of the resulted lattice model   to find
the interaction-dependent Wannier functions. Our results provide a general variational framework 
for transforming a periodic system into a lattice model self-consistently and can be generalized 
to systems where multiple bands are needed.

One can reformulate the nonlinear equation for Wannier functions in terms of Bloch functions,
and obtain a set of nonlinear equations for the Bloch functions. Simplified forms for these
nonlinear equations are obtained for two special cases, the superfluid regime and deep in 
the Mott insulator regime.  Our results are illustrated with a system of
 double-well potential, where some general properties of the nearest neighbor tunneling 
parameter $J$ and the on-site interaction $U$ are revealed.

Although our approach can be applied to other systems, e.g.,
fermionic systems, we here focus on the system of ultracold bosons in optical 
lattices\cite{Jaksch1998PRL,Morsch2006RMP,Yukalov2009LP}.
For bosons of mass $m$, the Hamiltonian  is
\ba
\hat{H}&&=\int d\r\,\hat{\psi}^\dagger({\r})\Big[-\frac{\hbar^2}{2m}\nabla^2
+V(\r)\Big]\hat{\psi}({\r})+\nonumber\\
&&\frac{1}{2}\int d\r d\r^\prime\Big[\hat{\psi}^\dagger(\r)\hat{\psi}^\dagger(\r^\prime)
U(|\r-\r^\prime|)\hat{\psi}(\r)\hat{\psi}(\r^\prime)\Big]\,,
\ea
where $v(\r)$ is a periodic potential and $u(|\r|)$ is the interaction between two atoms.  
If one is interested only in the low temperature properties of  the system, 
one can  use the single-band approximation and expand the bosonic field operator as 
\be
\hat{\psi}(\r)=\sum_j\hat{a}_jW_j(\r)\,,
\label{eq:wan}
\ee
where $W_j(\r)=W(\r-\r_j)$ is the Wannier function at site $j$ and $\hat{a}_j$ is the associated
annihilation operator. In this case, the trial ground state $|G_t\rangle$ is given by
$
|G_t\rangle=F(\hat{a}_j^\dagger)|\textrm{vaccum}\rangle\,,
$
where the functional $F$ is to be determined. Usually, the Wannier function in Eq.(\ref{eq:wan})
is pre-determined.  Here  the Wannier function is not known {\it a priori} 
except that it is expected to resemble the single-particle Wannier function.  We look for the
Wannier functions that minimize the system's  single-band ground
state $|G_t\rangle$,
\be
\label{eg}
E_G=\langle G_t|\hat{H}|G_t\rangle=\langle G_t|\hat{H}_{bh}|G_t\rangle\,,
\ee
where $\hat{H}_{bh}$ is the usual Bose-Hubbard Hamiltonian
\be
\label{bh1}
\hat{H}_{bh}=-\sum_{j_1j_2}J_{j_1j_2}\hat{a}^\dagger_{j_1}\hat{a}_{j_2}+
\sum_{j_1j_2}^{j_3j_4}U_{j_1j_2j_3j_4}\hat{a}^\dagger_{j_1}
\hat{a}^\dagger_{j_2}\hat{a}_{j_3}\hat{a}_{j_4}\,.
\ee
The parameters are given by
\ba
J_{j_1j_2}&=&-\int d\r W^*_{j_1}(\r)H_0 W_{j_2}(\r)\,,\\
U_{j_1j_2j_3j_4}&=&\frac{1}{2}\int d\r d\r^\prime \Big[W^*_{j_1}(\r)W^*_{j_2}(\r^\prime)\nonumber\\
&&\times U(|\r-\r^\prime|)
W_{j_3}(\r)W_{j_4}(\r^\prime)\Big]\,,
\ea
where
$
H_0=-\frac{\hbar^2}{2m}\nabla^2+V(\r)\,.
$
Usually further approximation is made so that only  two terms, the nearest 
neighbor tunneling  and the on-site interaction, are kept in the Bose-Hubbard model.  
For the purpose of deriving general formalism, this approximation is not needed so far.

We achieve the minimization of  the ground state energy $E_G$ by varying the Wannier function
under the  orthonormal constraints
$
h_{j}=\int d\r W^*(\r)W(\r-\r_j)=\delta_{0,j}\,.
$
According to the Feynman-Hellman theorem, we have 

\ba
&&\frac{\delta (E_G-\sum_j\mu_j h_j)}{\delta W^*(\r)}\nonumber\\
&=&\langle G_t|\frac{\delta \hat{H}_{bh}}{\delta W^*(\r)}|G_t\rangle
-\sum_j\frac{\mu_j\delta h_j}{\delta W^*(\r)}=0\,,
\ea
where $\mu$'s are Lagrangian multipliers.  
we obtain a  nonlinear equation for the Wannier functions
\ba
&&\sum_j\mu_jW(\r-\r_j)\nonumber\\
&=&
\sum_{j_1,j_2}\langle\hat{a}^\dagger_{j_1}\hat{a}_{j_2}\rangle
H_0 W(\r+\r_{j_1}-\r_{j_2})\nonumber\\
&+&
\sum_{j_1j_2}^{j_3j_4}\langle\hat{a}^\dagger_{j_1}
\hat{a}^\dagger_{j_2}\hat{a}_{j_3}\hat{a}_{j_4}\rangle\int d\r^\prime
\Big[ W^*(\r^\prime+\r_{j_2}-\r_{j_1})\nonumber\\
&\times& W(\r^\prime+\r_{j_2}-\r_{j_3})U(|\r^\prime-\r|)\Big]W(\r+\r_{j_2}-\r_{j_4})\,.\nonumber\\
\label{weq}
\ea
It is clear that the above equation depends on the ground state $|G_t\rangle$ of
the Bose-Hubbard model in Eq.(\ref{bh1}), which can be found with many-body 
methods  such as direct diagonalization,
Gutzwiller projection\cite{Rokhsar1991PRB},  density matrix renormalization group (DMRG)\cite{White1992PRL}, 
or time-evolving block decimation (TEBD)\cite{Vidal2004PRL}. At the same time, we need the Wannier functions to compute
$J$'s and $U$'s for the Bose-Hubbard model. Therefore, the Bose-Hubbard model in Eq.(\ref{bh1})
has to be solved self-consistently with the above nonlinear equation. We call this  
self-consistent single-band approximation.  
Due to the complexity of 
the equations, one can apply further approximations. For example, 
one can opt to completely ignore the interaction while solving Eq.(\ref{weq}).
This is just what people have traditionally done with the single-band approximation. 
One can also choose to keep only the nearest neighbor tunneling $J$
and the on-site interaction $U$ in the Bose-Hubbard Hamiltonian (\ref{bh1}). However,
the off-site terms  in the last summation in Eq.(\ref{weq}) can not be dropped simultaneously.

A periodic system can be described alternatively with Bloch functions.  
If we place the system in a box of $N$ lattice sites,  the Wannier functions are related
to Bloch functions as
\be
W(\r-\r_n)=\frac{1}{\sqrt{N}}\sum_{\k} e^{-i\k\cdot\r_n}\Psi_{\k}(\r)\,.
\ee
where $\Psi_{\k}$ is a Bloch function with Bloch wave number $\k$ and is normalized to one. 
For Bloch functions, the nonlinear equation (\ref{weq}) becomes
\ba
&&\sum_{\k}\nu_{\k}\Psi_{\k}(\r)=\sum_{\k}
\langle\hat{b}^\dagger_{\k}\hat{b}_{\k}\rangle H_0\Psi_{\k}+ %\frac{1}{N}
\sum^{<\k_1\k_2}_{\k_3\k_4>}\langle\hat{b}^\dagger_{\k_1}
\hat{b}^\dagger_{\k_2}\hat{b}_{\k_3}\hat{b}_{\k_4}\rangle\nonumber\\
&&\times\int d\r^\prime\Big[\Psi^*_{\k_1}(\r^\prime)\Psi_{\k_3}(\r^\prime)U(|\r^\prime-\r|)\Big]\Psi_{\k_4}(\r)\,,
\ea
where  $<\!\k_1\k_2\k_3\k_4\!>$  stands for summation  with the constraint $\k_1+\k_2=\k_3+\k_4+\textbf{K}$,
$
\hat{b}_\k=\frac{1}{\sqrt{N}}\sum_n\hat{a}_n e^{-i\k\cdot\r_n}\,,
$
and
$
\nu_\k=\frac{1}{N}\sum_n\mu_n e^{-i\k\cdot\r_n}\,.
$
$\textbf{K}$ is a reciprocal lattice. 
While we can not split Eq.(\ref{weq}) into a set of equations for Wannier functions at different sites, 
we are allowed to split the above equation for different Bloch wave numbers $\k$ and obtain
\ba
\tilde{\nu}_{\k}\Psi_{\k}(\r)&=& H_0\Psi_{\k}(\r)+ %\frac{1}{N}
\sum^{<\k_1\k}_{\k_3\k_4>}P_{\k_1\k\k_3\k_4}
\int d\r^\prime\Big[\Psi^*_{\k_1}(\r^\prime)\nonumber\\
&&\Psi_{\k_3}(\r^\prime)\times U(|\r^\prime-\r|)\Big]\Psi_{\k_4}(\r)\,,
\label{bloch}
\ea
where  $P_{\k_1\k \k_3\k_4}=\langle\hat{b}^\dagger_{\k_1}
\hat{b}^\dagger_{\k}\hat{b}_{\k_3}\hat{b}_{\k_4}\rangle/\langle \hat{b}^\dagger_{\k}\hat{b}_{\k}\rangle$
and $\tilde{\nu}_\k=\nu_\k/\langle \hat{b}^\dagger_{\k}\hat{b}_{\k}\rangle$.
In the following discussion, for simplicity, we shall use dilute atomic gases, where $u(|\r|)=g_0\delta(\r)$, %with $g_0=4\pi\hbar^2 a_s/m$ 
\cite{Pethick2002Book} to discuss two special cases.

We consider first the superfluid regime. With the Bogoliubov mean-field theory \cite{Pethick2002Book}, 
we have 
\ba
\hat{H}_{bh}&=&\epsilon_0{{\cal N}_0}+U_0{\cal N}_0^2+
\sum_{\k\neq 0} \Big[\epsilon_\k+4{\cal N}_0U_\k\Big]\hat{b}^\dagger_\k\hat{b}_\k+\nonumber\\
&&\sum_{\k\neq 0}{\cal N}_0U_{\k}
\Big(\hat{b}^\dagger_{\k}\hat{b}^\dagger_{-\k}+\hat{b}_{\k}\hat{b}_{-\k}\Big)\,,
\ea
where ${\cal N}_0$ is the number of atoms in the state $\psi_0$, $\epsilon_\k=\int d\r\psi^*_\k H_0\psi_\k$, and
$U_\k=(g_0/2)\int d\r |\psi_\k|^2|\psi_0|^2$.  Following the standard procedure \cite{Pethick2002Book},  we 
find that 
\ba
&&P_{\k_1\k \k_3\k_4}=v_{k_1}^2(\delta_{{\k_1},\k_4}\delta_{{\k},\k_3}+\delta_{{\k_1},\k_3}\delta_{{\k},\k_4})\nonumber\\
&&+\frac{u_{k}^*u_{k_3}v_{k_3}}{v_{k}}\delta_{{-\k_1},\k}\delta_{{-\k_3},\k_4}
-2\delta_{\k_1,0}\delta_{\k,0}{\cal N}_0\,,
\ea
where $u_0=v_0=\sqrt{{\cal N}_0}$,  $v_{\k\neq 0}^2=[(\epsilon_\k+4{\cal N}_0U_\k)/{\cal E}_\k-1]/2$, 
and $u_{\k\neq 0}^2=1+v_{\k\neq 0}^2$. ${\cal E}_\k=\sqrt{(\epsilon_\k+4{\cal N}_0U_\k)^2-4{\cal N}_0^2U^2_\k}$. 
This leads to a set of simplified nonlinear equations for Bloch functions
\be
\tilde{\nu}_{0}\Psi_{0}=H_0\Psi_{0}+g_0{\cal N}_0|\Psi_{0}|^2\Psi_{0}+
g_0\sum_{\k\neq 0}v_{\k}^2|\Psi_{\k}|^2\Psi_{0}\,,
\label{gp2}
\ee
and for $\k\neq 0$
\ba
\tilde{\nu}_{\k}\Psi_{\k}&=&H_0\Psi_{\k}+ 2g_0{\cal N}_0|\Psi_{0}|^2\Psi_{\k}+\nonumber\\
&&g_0\sum_{\k^\prime\neq 0}\Big[u_{k^\prime}v_{k^\prime}\frac{u_k}{v_k}+2v_{k^\prime}^2
\Big]|\Psi_{\k^\prime}|^2\Psi_{\k}\,.
\ea
Since $u_k$ and $v_k$ themselves depend on $\psi_{\k}$, the above two equations have to be solved
self-consistently. Note that in the above derivation we have assumed that
the lattice potential is symmetric, $V(\r)=V(-\r)$, so that $\psi_{-\k}=\psi^*_{\k}$. We have also ignored
the scattering processes with non-zero $\textbf{K}$. 

In contrast to superfluid regime, deep in the Mott-insulator regime,  we  have 
\ba
\langle\hat{a}^\dagger_{j_1}\hat{a}_{j_2}\rangle&=&n_0\delta_{j_1,j_2}\,,\\
\langle\hat{a}^\dagger_{j_1}
\hat{a}^\dagger_{j_2}\hat{a}_{j_3}\hat{a}_{j_4}\rangle&=&n_0^2\delta_{j_1,j_3}\delta_{j_2,j_4}
+n_0^2\delta_{j_1,j_4}\delta_{j_2,j_3}\nonumber\\
&&-(n_0^2+n_0)\delta_{j_1,j_2}\delta_{j_2,j_3}\delta_{j_3,j_4}\,.
\ea
where $n_0$ is the averaged number of bosons per site. 
In this case, Eq.(\ref{weq}) is simplified and has the form
\ba
\frac{\mu_0}{N_0}W(\r)&=&H_0W(\r)+g_0n_0\sum_{\r_j\neq 0}|W(\r-\r_j)|^2W(\r)\nonumber\\
&&+g_0(n_0-1)|W(\r)|^2W(\r)\,.
\label{gp3}
\ea
The off-site terms on the right hand side is usually very small and can be ignored. 
With this in mind, we immediately have one observations. With one atom per site, i.e.,
$n_0=1$, the Wannier function is approximately the single-particle ground state wave function of 
each individual well of the lattice; there is  no interaction effect on the Wannier function.

We now use one dimensional double-well potential with periodic boundary condition 
to illustrate our theory.  The two Wannier functions for the left well and the right well are related to the
ground state and the first excited state for the double-well potential as follows
\be
W_l=\frac{\sqrt{2}}{2}(\Psi_0+\Psi_1)\,,
~~~~W_r=\frac{\sqrt{2}}{2}(\Psi_0-\Psi_1)\,,
\ee
where $\Psi_0$ and $\Psi_1$  are chosen so that they are both positive in the left well. 
These two Wannier functions satisfy the nonlinear
equation (\ref{weq}).
The corresponding Bose-Hubbard model is
\ba
\hat{H}_{2}&=&\Big[-J+2(N_0-1)U_3\Big](\hat{a}^\dagger_{l}\hat{a}_{r}+\hat{a}^\dagger_{r}\hat{a}_{l})+\nonumber\\
&+&U_2(\hat{a}^\dagger_{r}\hat{a}^\dagger_{r}\hat{a}_{l}\hat{a}_{l}+
4\hat{a}^\dagger_{l}\hat{a}_{l}\hat{a}^\dagger_{r}\hat{a}_{r}+
\hat{a}^\dagger_{l}\hat{a}^\dagger_{l}\hat{a}_{r}\hat{a}_{r})\nonumber\\
&+&U(\hat{a}^\dagger_{l}\hat{a}^\dagger_{l}\hat{a}_{l}\hat{a}_{l}+
\hat{a}^\dagger_{r}\hat{a}^\dagger_{r}\hat{a}_{r}\hat{a}_{r})\,,
\label{bh2site}
\ea
where $U_2=U_{llrr}$, $U_3=U_{lrrr}$, and $U=U_{llll}$. 

%%%%%%%%%%%%%%%%%%%%%%%%%%%%%%%%%%
\begin{figure}[!tbp]
\includegraphics[width=0.7\linewidth,angle=0]{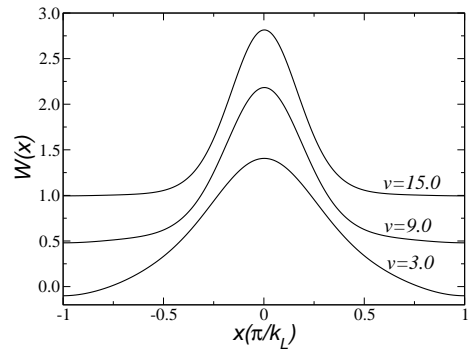}
\caption{Wannier functions for different depths of the double-well potential. 
For clarity, the curves for $v=9.0$ and $v=15.0$ are shifted up by 0.5 and 1.0, respectively. 
$c=2.5$. The unit of $v$ is the recoil energy $E_r=\hbar^2k_L^2/2m$.}
\label{wan}
\end{figure}
%%%%%%%%%%%%%%%%%%%%%%%%%%%%%%%%%
We choose the double-well potential as a part of the one dimensional optical lattice created experimentally
in Ref.\cite{Paredes2004Nature}. So, the double well potential is given by
$
V(x)=V_0\sin^2(k_Lx)\,,
$
where $k_L$ is the wave number of the laser that creates the potential. Due to the lateral confinement, 
the interaction strength $g_0$ is given by
$
g_0=(4\pi\hbar^2 a_s/m)(m\omega_{\perp}/(2\pi\hbar))=2\hbar\omega_\perp a_s\,,
$
where $a_s$ is the $s$-wave scattering length and $\omega_\perp$ the perpendicular confinement frequency.
We consider the case of $n_0=1(N_0=2)$. 
In our numerical calculations, the Hamiltonian in  Eq.(\ref{bh2site}) is diagonalized directly and Eq.(\ref{weq}) is
solved  with the nonlinear equation solver in MATLAB. 
And we use $c=\pi m g_0/(\hbar^2k_L)$ as the dimensionless interaction parameter.

The Wannier functions found numerically are plotted in Fig.\ref{wan}.  As expected, the Wannier function becomes
more and more localized as the depth of the well increases.  It is worthwhile to note that
the Wannier functions for shallow wells (e.g., $v=3.0$) have nodes, which shows that nodeless
ground state does not necessarily imply a nodeless Wannier function.

\begin{figure}[tbp]
\includegraphics[width=0.9\linewidth,angle=0]{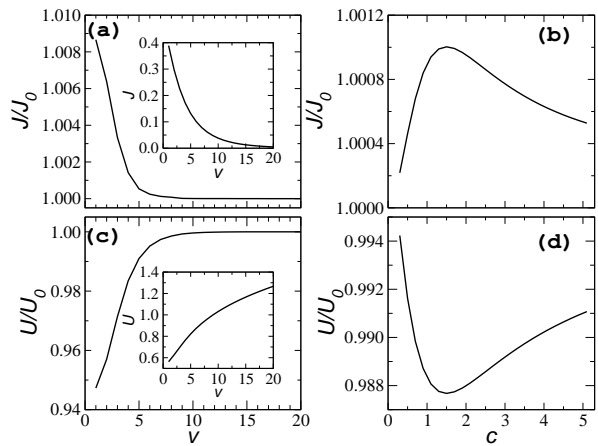}
\caption{The tunneling parameter $J$ and the on-site interaction $U_4$ as functions of the double-well depth. 
 For (a) and (c), $c=2.5$; for (b) and (d), $v=5.0$. The energy unit  is the recoil energy $E_r=\hbar^2k_L^2/2m$.}
\label{JU}
\end{figure}

The numerical results for the dependence of $J$ and $U$ on the well depth and interaction strength are shown 
in terms of the ratios $J/J_0$ and $U/U_0$ in Fig.\ref{JU}. $J_0$ and $U_0$ are the tunneling parameter and 
on-site interaction obtained with single-particle Wannier function. 
Certain interesting  behaviors of $J/J_0$ and $U/U_0$ are revealed. 
For a fixed interaction strength, both $J/J_0$ and $U/U_0$ approach one as the lattice gets stronger. 
If the lattice strength is fixed and the interaction strength changes, $J/J_0$ and $U/U_0$ reach 
their respective extremum values at a 
critical interaction strength.

%%%%%%%%%%%%%%%%%%%%%%%%%%%%%%%%%%
%%%%%%%%%%%%%%%%%%%%%%%%%%%%%%%%%

We argue that the behaviors of $J/J_0$ and $U/U_0$ revealed in Fig.\ref{JU} is not limited to the 
double-well systems, and should hold generally for ultracold bosons in optical lattices.  
As already discussed, deep in the Mott insulator regime, the solution of  Eq.(\ref{gp3}) is 
single-particle Wannier function for $n_0=1$.  This implies that both of the ratios $J/J_0$ and $U/U_0$
should approach one as the optical lattice gets stronger with fixed interaction strength. 
This is exactly what is seen in Fig. \ref{JU}(a,c). If one fixes the lattice strength and 
increases the repulsive interaction,  then one has single-particle Wannier function at
both the beginning ($c=0$) and the end ($c\gg 1$) of this change. This indicates that
$J/J_0$ should reach its maximum point and $U/U_0$ arrive at its minimum point during
this process as shown in Fig. \ref{JU}(b,d). If one simulates the superfluid-Mott insulator
transition with the self-consistent single band approximation proposed here,
the behaviors of $J/J_0$ and $U/U_0$ in Fig. \ref{JU}(b,d) can be used to determine
the transition point between superfluid and Mott insulator.

Note that the interaction effect on $J$ and $U$ is much smaller compared to the mean-field results in Ref.\cite{Liang2009}.
This is likely due to the finite size of our example system, where   the Wannier function is confined and can not extend
to infinity.  More extensive numerical studies are needed to clarify this. If future numerical computation
indeed indicates that the interaction effect on $J$ and $U$ is as large as indicated in 
the mean-field results \cite{Liang2009}, then the behaviors of $U/U_0$ in Fig. \ref{JU}(c,d) 
may be used as an experimental means to detect the superfluid-Mott insulator transition as
$U$ can now be measured experimentally \cite{Campbell08042006} and the interaction can be adjusted with the
Feshbach resonance\cite{Weiner1999RMP}.

Currently in typical experiments,  ultracold atoms are also trapped by a harmonic 
potential\cite{Morsch2006RMP,Yukalov2009LP}. Consequently, the wells are not identical to each other. 
Also a random potential can be added to make the wells non-identical.
Nevertheless, single-band approximation can  still  be applied  as long as the difference
between the site energies is smaller than the energy gaps between
the ground state and the first excited state in the wells.   For simplicity,
we consider a one dimensional potential of $N$ wells, which are not identical. 
In this case, the Wannier functions
for the lowest ``band" can be defined as
\be
W_j(x)=\frac{1}{\sqrt{N}}\sum_{k=0}^{N-1} e^{i\frac{2kj\pi}{N}}\psi_k\,,~~j=0,1,,\cdots, N-1\,,
\ee
where $\psi_k$'s are the lowest $N$ eigenstates.  
Our variational approach can be easily adopted to this case with just one 
modification. Since the Wannier functions at different sites have
different shapes, the constraint is now
$
h_{n,m}=\int dx W^*_n(x)W_m(x)=\delta_{n,m}\,.
$
As a result, one obtains a set of nonlinear equations for the Wannier functions.
Since everything is straightforward, we shall not write out the equations here. 

%In sum, we have presented a general theoretical framework for self-consistent single-band 
%approximation.% that converts a periodic system to a lattice model. 
%Within this framework, we have obtained a set of nonlinear equations for either Wannier 
%functions or Bloch functions, which have to be solved together self-consistently the ground state
%of the resulted lattice model. 
%This framework can be generalized to systems where multiple bands have to be considered. 

This work was  supported by the ``BaiRen'' program of the 
Chinese Academy of Sciences, the NSF of China (10604063,10825417) 
and the MOST of China (2005CB724500,2006CB921400). B.W. acknowledges
the hospitality of the Aspen Center for Physics.

%\bibliography{/Users/wubiao/Documents/references/general}
%\bibliographystyle{apsrev}

\end{document}